\documentclass[12pt,a4paper,onecolumn]{article}
\usepackage{url}
\usepackage[top=1.9cm,right=2.54cm,left=3cm,bottom=2.54cm]{geometry}
\usepackage{graphicx}

\title{A Participatory Sensing Framework for Environment Pollution Monitoring and Management
}
\author{Al Amin Neaz Ahmed$\dagger$, H.M Fazlul Haque$\dagger$, Abdur Rahman$\dagger$, \\Md Susam Ashraf$\dagger$, Sanjay Saha$\dagger$
and Swakkhar Shatabda$\dagger$\thanks{correspondance: swakkhar@cse.uiu.ac.bd}\\
$\dagger$ Department of Computer Science and Engineering, \\United International University, Dhaka, Bangladesh}
\date{}
\begin{document}
\maketitle

\begin{abstract}
Effective monitoring and management of environment pollution is key to the development of modern metropolitan cities. To sustain and to cope with the exponential growth of the cities with high industrialization, expert decision making is very essential in this process. A good governance system must be supported by an actively participating population. In participatory sensing, individuals and groups engages in the data collection actively and the helps the city governance to make proper decisions. In this paper, we propose a participatory sensing based three-tier framework to fight environment pollution in urban areas of Bangladesh. The framework includes an android application named `My City, My Environment', a server for storage and computation and also a web server for the authority to monitor and maintain environmental issues through expert decision making. We have already developed a prototype system and deployed it to a small scale and demonstrated the effectiveness of this framework.  

\end{abstract}
\section{Introduction}

Community participation enables citizens to become active entity in the decision making and action process \cite{world2002community}. It is built on collective awareness and active thought process of the population. Nobel Prize winner Elinor Ostrom in her `Governing the Commons' theory has emphasized on the support of users (citizens) for effective monitoring and enforcement of rules \cite{ostrom1990ostrom}. She also stressed on low cost process through which the use of resources by humans are to be monitored and information are verified. This have encouraged the modern authorities to design a active participatory framework for governance. Communities that maintain frequent face-to-face communication and dense social network and are engaged actively in the governance.

Participatory sensing \cite{burke2006participatory,campbell2006people} is the concept where individuals, groups and communities engages in the data collection actively to construct knowledge. This is mostly done in two ways: using a large sensor network and using the available devices (e.g. mobile phones) to build ad hoc networks. In this paper, we address the issue of participatory sensing in the governance context of Bangladesh.

Many developing countries like Bangladesh are going to a rapid phase of industrialization and urbanization. In the process of these, comes the issue of environment. Almost every direction in the development triggers some environmental issue. The explosion of urbane population adds to it and makes it more unmanageable. In metropolitan cities like Dhaka, the capital of Bangladesh, even garbage management are sometimes not done properly. Many bins are left full roadside and it brings terrible experience for the city-dwellers. There are other environmental issues like noise, air, water, visual, light etc. 

In the traditional way of management, except for scheduled work, there is a lack of proper communication for the incidents happening (e.g. environment pollution) and accountability after the measures are taken. Corrupt authority adds more to spoil the broth. In many cases, the use of social media or social network (e.g. Facebook, Twitter) has been successful in creating awareness among the community through active participation of the people in reporting and supporting the cause. A successful example of such is the BP oil spill incident in Mexican Gulf in 2010 \cite{starbird2015social}. International reports \cite{declaration1992rio} have also emphasized the necessity of active participation of all users to move towards sustainable development.

The solutions adopted in sensing and monitoring environment pollution employed in large cities involve high cost of development due to the required infrastructure. Even in the cases of participatory sensing or crowd-sourcing the systems often come with sophisticated sensor devices \cite{hasenfratz2012participatory,rana2010ear}. Crowd-sourcing and participatory sensing in the context of Bangladesh is rater underutilized. Some applications include fighting sexual harassment \cite{ahmed2014protibadi} and data collection through monetary incentive given to the participants \cite{bell2016real}.  
 
In this paper, we propose a participatory sensing framework for environment pollution monitoring and management in large metropolitan cities like Dhaka, capital of Bangladesh. We propose a three tier application framework that will ensure the participation of the responsible citizen through a mobile phone based application, a server that collects and analyzes data from end-users in real time and a web based application for the city governance to help with decision making  regarding the monitoring and management of environment pollution. 

Using the mobile phone based application, an end user will be able to report any nearby incident of environment pollution or hazard by taking a photo or video of the problem and submitting it to the server. Each report will be tagged using some meta information and location using positioning system provided by GPS of the mobile phone. The trust issues will be solved using social community based reporting and reliable user ratings. The system will also offer anonymity and ensure privacy of the end-users. In the medium tier, the server collecting the information will categorize and analyze the information posted by end-users automatically using intelligent algorithms. It will provide map-based meta information to the other two tiers of the framework. The end-users will be able to watch and support the reports submitted by other users. The mobile phone based application will also have a social network based interface to ensure the participation of other users through already established popular social network services like Facebook, Twitter and blogs. The application in the third tier will help the municipality governing authority to monitor and detect potential threats (short term and long-term) by visualizing them based on meta-information and community trust and the application will help them to take decisions based on an expert decision system. Our framework is supported by a feedback system by the end-users which will in effect increase the  accountability of the city governing body. We have developed a prototype for our proposed system and tested it with sample data and demonstrate the usability of the system.
\subsection{Related Work}
Collective participation of citizens is very essential in management and assessment of environmental issues \cite{declaration1992rio}. With the advent of modern technology and a global boom in the number of mobile devices (phones etc.) and low cost sensors participatory sensing \cite{burke2006participatory} and crowd-sourcing \cite{campbell2006people} are in application to solve the problem of effective 
data acquisition and assessment in fields like health, urban planning, environment, education etc. There has been a number of successful applications of crowd-sourcing all over the world ranging from collecting epidemiological data in England \cite{aanensen2009epicollect} to food security data in Africa \cite{enenkel2015food}.  

There are several challenges in crowd-sourcing mobile phone data in urban spaces related with incomplete samples, unknown contexts, user privacy,  trustworthiness of data and conservation of energy \cite{kanhere2011participatory}. Moreover, most of these applications comes with use of sensor devices, those though low cost comes with extra added expense in terms of money and maintenance.   

Among other fields participatory sensing and crowd-sourcing has been used effectively in monitoring road and traffic conditions \cite{mohan2008nericell}, crime watch \cite{williams2013crowdsourcing}, disaster management \cite{horita2013use}, disease spread \cite{chen2016syndromic}. Even scientific problems like optical character recognition \cite{narula2011mobileworks} and protein structure prediction \cite{khatib2011crystal} are solved effectively by crowd-sourcing and participatory sensing applications. 

Environment pollution is an area where participatory sensing can play a lead role in data collection and management \cite{kanhere2011participatory}. Effective waste management in urban area is handled in  \cite{offenhuber2012putting,kubasek2013crowdsource}. The air quality control or monitoring applications often comes with low cost sensors interfaced with mobile phones or devices. Such examples are  Gas-Mobile \cite{hasenfratz2012participatory}, Common Sense \cite{dutta2009common} and P-sense \cite{mendez2011p}. An air temperature monitoring system was proposed in  \cite{overeem2013crowdsourcing} that collects air temperature from the temperature of the mobile phone batteries. A successful application of crowd-sourcing and participatory sensing in water monitoring in World Water monitoring challenge \footnote{http://www.monitorwater.org/}. It includes 1458517 participants from 142 countries in the world reporting 70,927 water bodies \cite{wwmc}. Creek Watch \cite{kim2011creek} uses iPhone application to employ users to aid water management. Iphone camera is also used in reporting water quality in \cite{leeuw2014crowdsourcing}. 

An WebGIS framework is used to monitor and manage soil and water in \cite{werts2012integrated}. NextDrop \cite{kumpel2012nextdrop} is used to track water distribution and supply in Indian communities. Ear-Phone \cite{rana2010ear} was proposed a low cost solution to urban noise mapping through participatory sensing. NoiseTube \cite{stevens2010crowdsourcing}, a mobile based sensing application and a web-based community memory was deployed as a proposed solution to monitor noise pollution in ubiquitous urban setting. London heatmap was generated using crowd-sourcing in \cite{chapman2016can}. Among other successful applications are light pollution \cite{hintz2014thelma} and oil reporter \cite{starbird2015social}. The applications in Bangladesh are under-utilized \cite{ahmed2014protibadi} and sometimes involves monetary incentive given to the participants \cite{bell2016real}. Our approach is a cost-effective one since it does not use extra sensors and only based on the mobile phone devices and include the effective social network interaction for effective management.

\subsection{Paper Organization}
Rest of the paper is organized as follows: Section~\ref{secMethods} describes our methodology and details of the framework; in Section~\ref{secResults}, we present the implementation and discuss the findings and results that we gathered from the application that we developed and Section~\ref{secCon} concludes the paper with an indication of possible future work and delineating the limitations of the current system.
\section{Methodology\label{secMethods}}
In this section, we describe the methodology of the framework that we have proposed and developed in this research.

\subsection{Architecture}

\begin{figure}
\centering
\begin{tabular}{cc}
\includegraphics[scale=0.3]{three-tiers}& \includegraphics[scale=0.45]{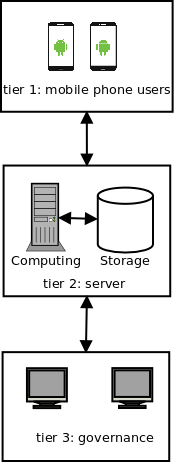}\\

\end{tabular}
\caption{Illustration of the three tiers of the proposed framework.}
\label{fig:three-tiers}
\end{figure}

This section demonstrates the system design of our proposed participatory sensing framework for environmental pollution monitoring and management. The framework is basically composed of three tiers as shown in Figure \ref{fig:three-tiers}. The first tier is a mobile application built for Android based mobile phones which will be used for collecting reports from end users. These report are stored and analyzed in the second tier. The third tier provides municipality authorities with regular reports on pollution and a map to pinpoint the location of the reports from the end users. 

\subsection{Smart Phone Application}

A mobile application name \textit{My City My Environment} (available at: \url{www.kipailam.com/app/mcme.apk}) is built to provide end users a responsive interface so that they can submit incidents of reports on environmental pollution and provide user rating on existing reports. The application is built using JAVA language for smart-phones running on Android OS. Following is the list of resources whose permissions are required for installation of the application:
\begin{itemize}
	\item \textbf{Camera}: Used for taking photos and videos to attach with submitted reports. 
	\item \textbf{Microphone}: Video recording requires permission for accessing microphone of the phone.
	\item \textbf{Location}: Required for adding GPS based location information along with pollution reports.
	\item \textbf{Storage}: This resource was required for accessing files from phone's storage to bind evidences of pollution.
\end{itemize}

\begin{figure}[h]
\centering
\begin{tabular}{ccc}
	\includegraphics[scale=0.13]{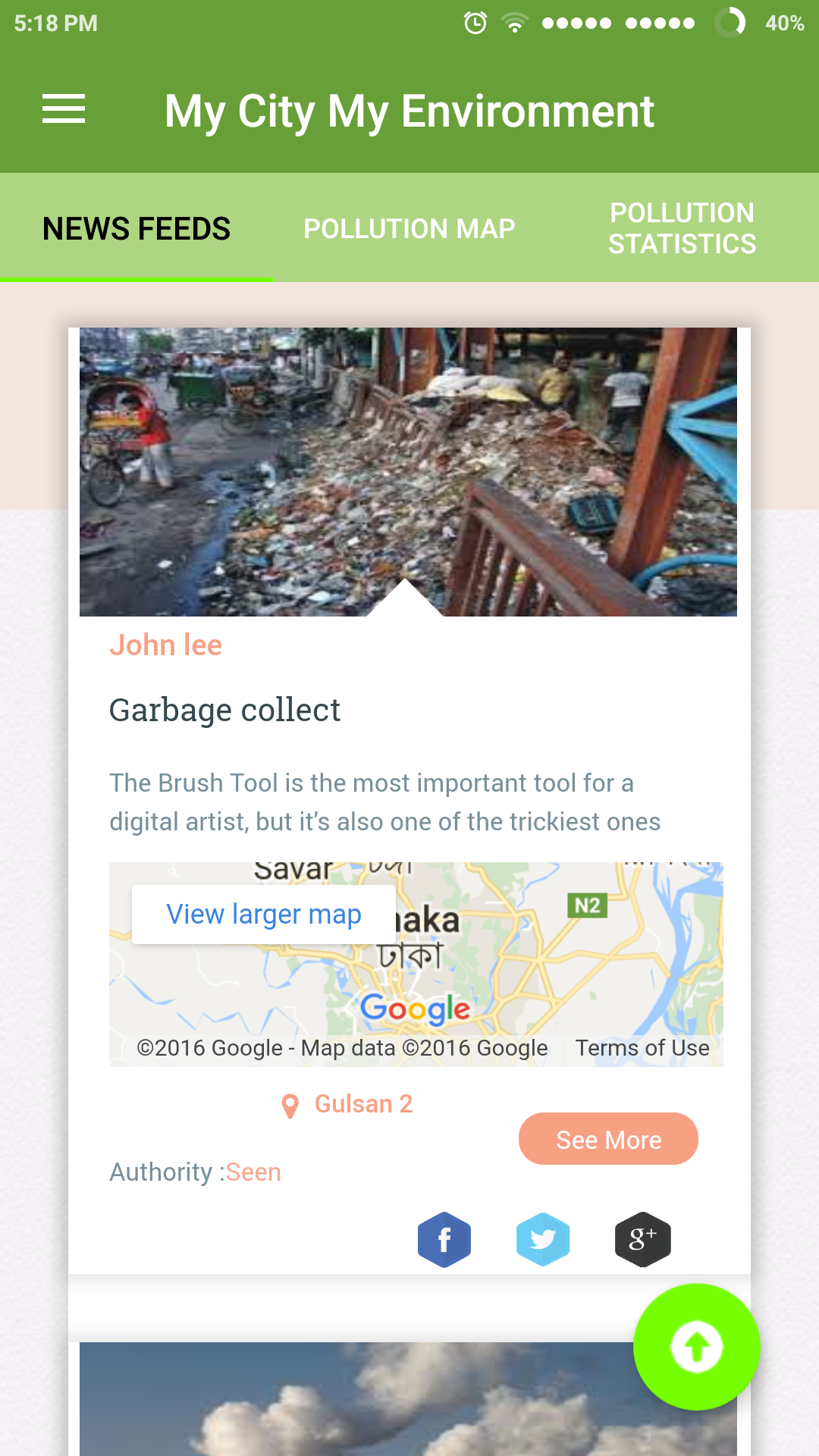}&
	\includegraphics[scale=0.13]{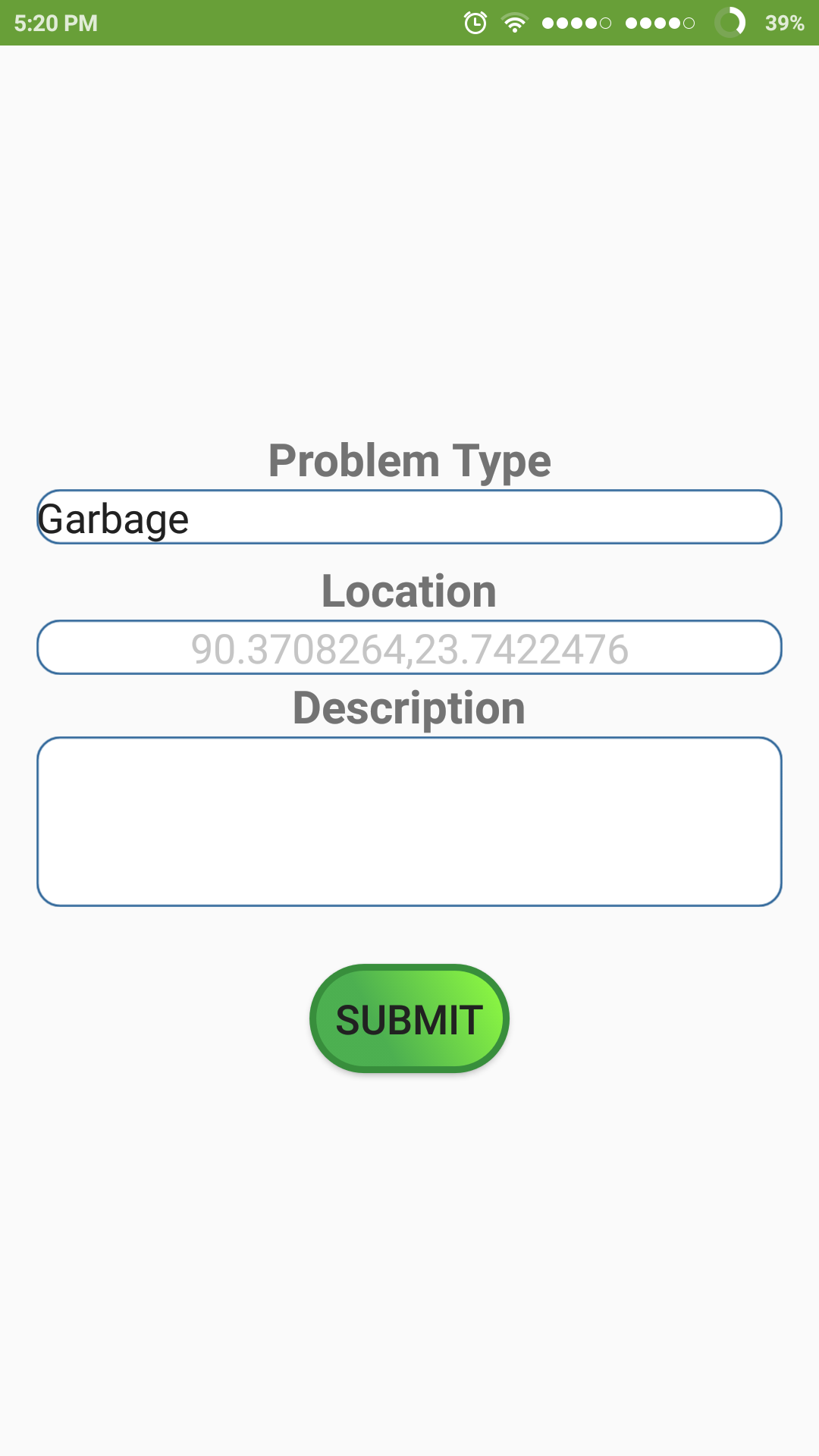}&
	\includegraphics[scale=0.13]{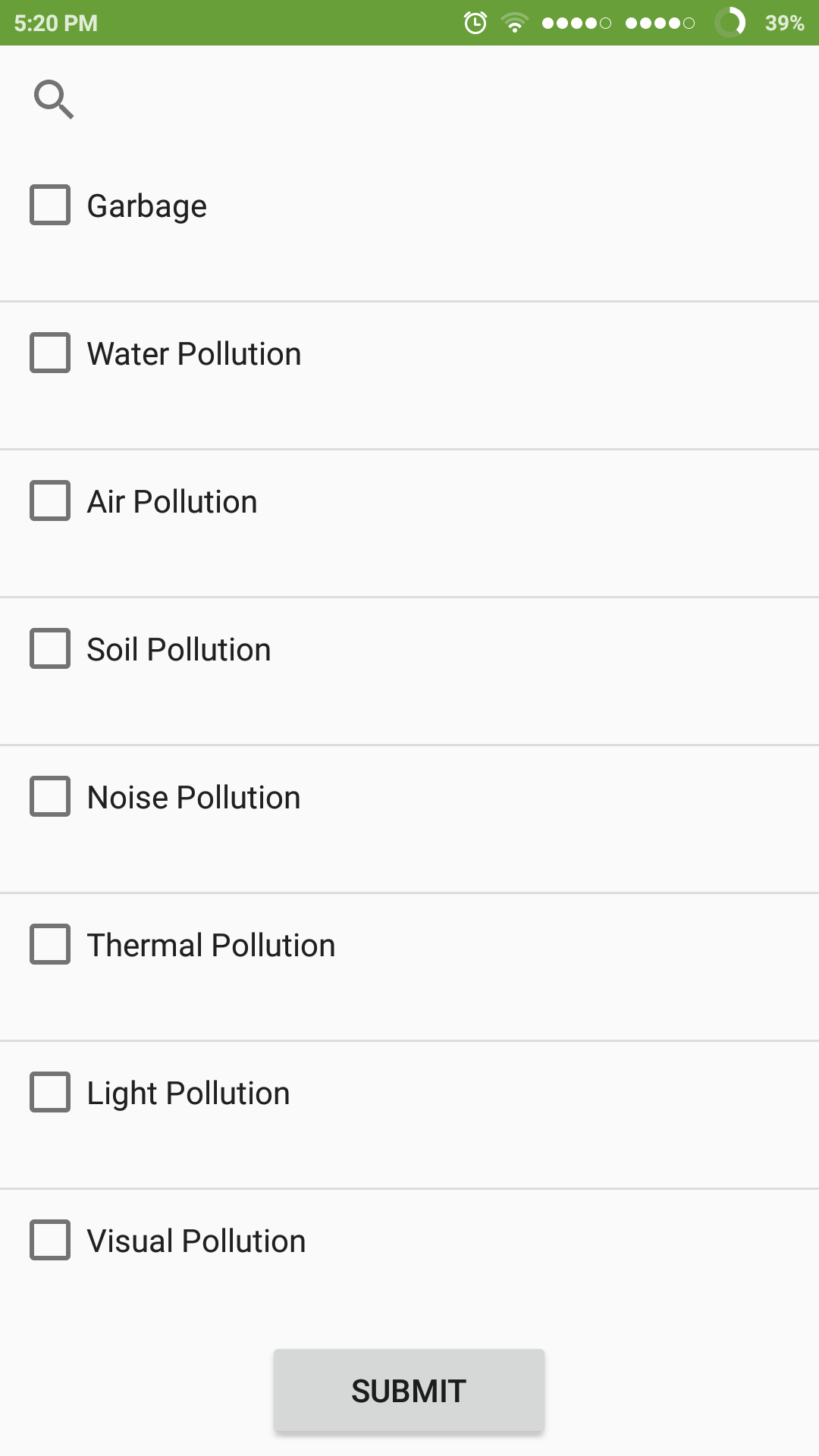}\\
	(a)&(b)&(c)\\
\end{tabular}

\caption{(a) News feed page shows the recent post on hazardous incidents of environment pollution (b) submission and (c) selection of pollution type in the application.}
\label{figScreen}
\end{figure}

The phone has an immensely user friendly interface to communicate with the framework. Users are required to login before they use the application. They will find a homepage which is a news feed showing recent posts on environment pollution submitted by other users. Figure \ref{figScreen}(a) shows how the home page may look like. This page contains a floating button on the bottom right corner which gives users with options to submit a report on environmental pollution.

Apart from taking photos or videos from inside the application, users need to provide a \textit{type} for categorizing the report for future usage (See Figure~\ref{figScreen} (b)). A list of possible \textit{types} will be presented as option for the users where they can select multiple \textit{types} to address any incident of environmental pollution. The mobile application collects the location information from available wifi network, mobile network or GPS (See Figure~\ref{figScreen} (c)). Recorded videos and images taken from within the application is stored in a different directory in the system storage. 

\begin{figure*}
\centering

\begin{tabular}{cc}
\includegraphics[scale=0.14]{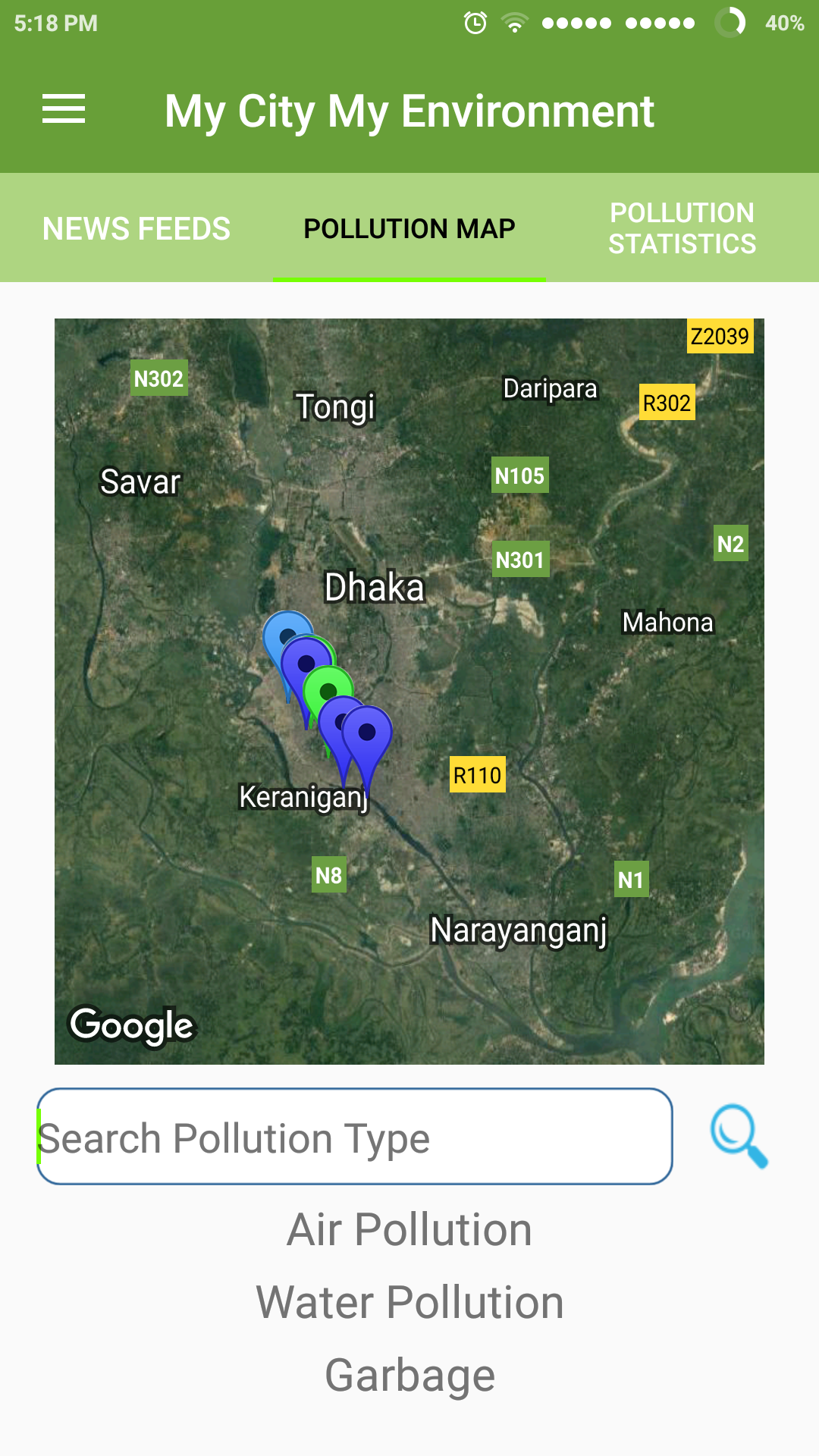}&
\includegraphics[width=0.65\linewidth]{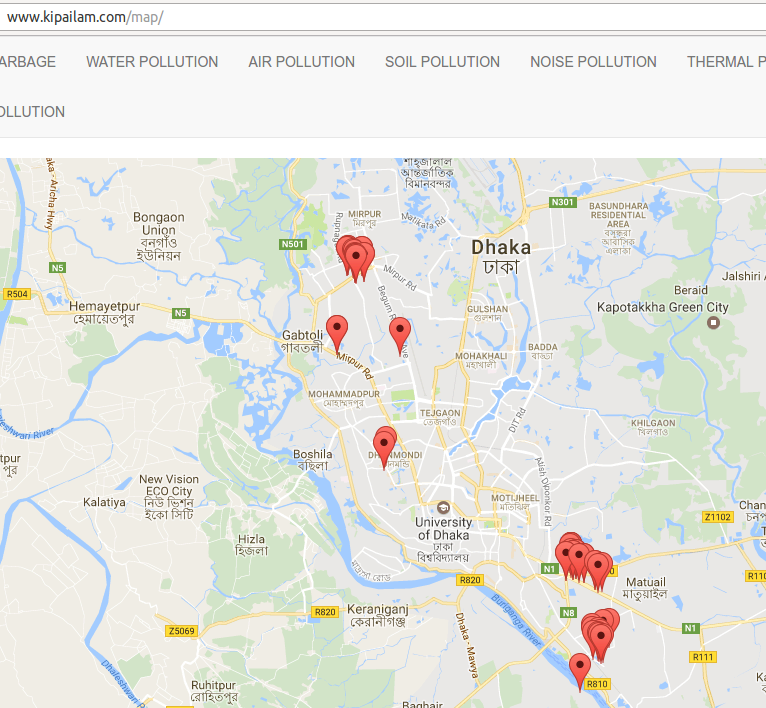}\\
(a)&(b)\\
\end{tabular}

\caption{Pollution map for (a) the users in the mobile application and (b) in the web application for the users in the third tier of the application.}
\label{fig:pollution-map}
\end{figure*}

Users can view the pollution map from the mobile application (See Figure~\ref{fig:pollution-map}(a)) which is eventually a map of geographic locations extracted from the reports submitted by the end users. Aside pollution map, users can also view statistics of environment pollutions of the city through the \textit{Pollution Statistics} tab.  

\subsection{Server Application} 

Storage of data collected from the end users and the web application are hosted in the server side which is the middle tier of our proposed framework. The server provides an Application Program Interface (API) for the mobile application and the web application; thus it is a facilitator of information for the other tiers. A secured, fast and sustainable architecture is used to build the middle tier. Following is the list of technological stack used for developing the server side:
\begin{itemize}
	\item Apache Web Server
	\item Laravel 5.3 Web framework for API and Web support (Language: PHP)
	\item JSON based API support
\end{itemize}

\subsection{Web Application}
Users in the third tier of our framework are the municipality authorities who will be able to see the pollution map as shown in Figure~\ref{fig:pollution-map}(b) through the web application (\url{www.kipailam.com/}) . The map is updated in real time, thus as soon as an end user's report is validated by the system, the map gets updated. Along with that, users of this web application can get (print or email) both detailed or summarized reports.
\section{Results\label{secResults}}
In this section, we delineate the results and findings that were gathered after the deployment of the framework in test basis. 

\subsection{Study Area}
To test the effectiveness of our framework and application, we designed a study. In our study, we used Dhaka, the capital city of Bangladesh. A group of volunteers were selected randomly. Most of them were students studying in the tertiary level or university. They usually travel in various directions in the city around their educational institutes, home and other places of recreation. All of these students were in possession of a smart phone equipped with gps, camera, etc that is the minimum requirement supported by our system. For the systems without internet connection, we provided an off-line submission procedure to the server of their data locally saved in the storage of the smart phone. The users were told to report any incident of environment pollution like: waste, air, water, noise, light, etc. 
\subsection{Deployment Period}
We told the participants to use both the `My City, My Environment' application and the web-server using a browser whenever they find any incident over a period of two weeks in December 2016. During this time, they reported 53 data points mostly in Dhanmondi, Mirpur and Jatrabari area. They shared several of these items in social media like twitter and Facebook.

\subsection{Responses}
\begin{figure}[h]
\centering
\includegraphics[scale=0.5]{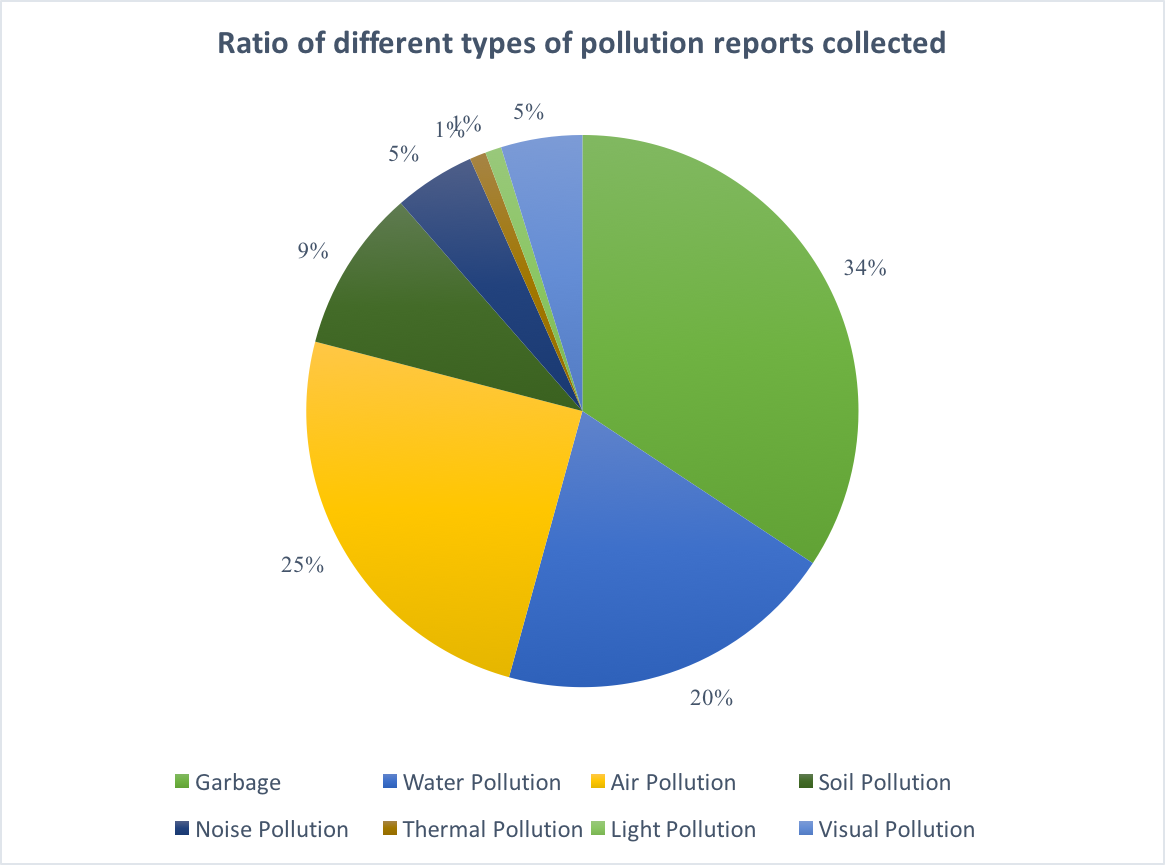}
\caption{Distribution of data among various types of pollutions reported. \label{piechart}}
\end{figure}

Various incidents of pollution were reported. Figure ~\ref{piechart} shows are distribution of different types of pollutions reported by the users through our application. Maximum was related to garbage management system (34\%). 

Many of the posts were supported by free text comments. A few examples are like '\textit{the canal is blocked by garbage}',  '\textit{this canal is totally blocked which causes water logging in the area during the rainy season}', '\textit{air pollutes because of these fish market}', '\textit{too much dust}', '\textit{road is covered by garbage}', '\textit{due to water logging for a long period}', '\textit{because of water logging, transportation is disturbed}', '\textit{drainage system is blocked because of garbage}', etc. 

\subsection{User Feedbacks}
After the deployment period, we engaged the users in a post-study feedback survey. This feedback also included a usability test for the UI of the application. The questionnaire contained five positive statements about the applications as following:
\begin{enumerate}
\item Learning to use the app was easy for me.
\item The app will be able to bring community awareness on environment pollution.
\item The app will be able to bring effective changes in decision making.
\item I will definitely try to use the app in future.
\item I will suggest it to my friends and colleagues. 
\end{enumerate}

The users were asked to rate this statements stating whether they `Strongly Agree', 'Agree' or 'Disagree' with the statements. Also we asked them to comment on the most positive and negative thing they found in the application. Twenty one users provided us with feedback and Table~\ref{tabUser} summarizes the user feedbacks from the questionnaire:

\begin{table}
\centering
\caption{Summarized results from user feedback survey \label{tabUser}}
\begin{tabular}{|c|c|c|c|}
\hline
Question no&Strongly Agree& Agree & Disagree\\
\hline
1&52.4\%&47.6\%&--\\
\hline
2&61.9\%&38.1\%&--\\
\hline
3&66.7\%&28.6\%&4.8\%\\
\hline
4&61.9\%&23.8\%&14.3\%\\
\hline
5&52.4\%&38.5\%&9.5\%\\
\hline
\end{tabular}
\end{table}

We could notice that mostly the users were in agreement with the statements except for a few disagreements. Among the positive things, the users told us about `\textit{Community Awareness through social media}', `\textit{one can play a very important role in this respect as smart phones are available to people of all classes nowadays}', `\textit{get to show what is going on with the city by this android app}'. Users emphasized about the requirement of `\textit{clean city}'and how he is able to `\textit{voice my opinions about environment pollutions}'. Among the negative feedbacks, they complained about the availability of internet, cost of bandwidth to upload images and lack of the third tier involvement in the application. There were feedback to include to tackle sexual or any type of harassment through the same framework.
\subsection{Challenges}
In this sub-section, we describe a few challenges to this framework that we faced during implementation, might be posed in future and how we address them.
\subsubsection{User Trust}
One of the strength of our framework is the openness that allows any user to submit data. However, there is a hinge to data trust worthiness. In present, we have registration system by which a user can register in the server and submit the reports using registration credentials. The similar reported rated high/positive by the other users will increase the trustworthiness of the particular user and incident and will resolve the issue. However, in future we plan to introduce a new layer of users with admin capabilities or high reputation who will verify and rate the incidents reported by the normal users. 
\subsubsection{User Privacy}
If not provided with privacy protection of the users, the application users sometimes might feel unsafe to report incidents. To address this issue within our framework, we also allow anonymous reports to be submitted and later supported and rated by the other users. 
\subsubsection{Cost Effectiveness}
The number of mobile phone users in Bangladesh have increased in a large extent \cite{brtc} and most of them are using android based very cheap smart phone sets. Thus our framework does not incur any extra cost in terms of engaging the users. There has been direct concern from the users of the app towards the cost of the application due to the connectivity needed to upload images and geo-locations. We have already addressed this issue by making the image/video an optional feature and also added off-line reporting system, where they will submit the data stored locally once they are connected. However, governance might address the issue by adding this to available free internet \cite{freebasic} or other such publicly available WiFi networks.
\subsubsection{User Incentives}
The idea of collective governance \cite{ostrom1990ostrom} would not come to reality unless we are successful in bringing a large number of regular active users in our community. To solve this issue, we have included the social media and network features. The users are motivated when their supported or reported cause or environmental issues get noticed, liked and shared by other users and addressed by the authority. In future, we wish to add gamification in smaller level to give them awards and thus incentives to use the application regularly.  
\section{Conclusion\label{secCon}}
In this paper, we described a participatory sensing framework for environment pollution monitoring and management. Our app `My City, My Environment' successfully engaged users to report and spread awareness among the citizens by sharing them in social network. The active governance feature is absent and generates a report and a visual map to help in expert decision making. However, these are the few areas in which we wish to explore and develop further:
\begin{itemize}
\item [--] Engage city governing bodies in the third tier of the application for a locality.
\item [--] Provide intelligent application and algorithms in the second tier to help the authority in solving problems more effectively. One such idea is to provide a vehicle routing solution to waste management \cite{benjamin2010metaheuristics}. 
\item [--] Add new features to improve the data trustworthiness and provide better incentives to the users through gamification of rewarding scheme. 
\end{itemize} 

We believe incorporating this features and active involvement of stakeholders in the framework is capable of producing effective results in monitoring and management of environment pollution in urban areas of Bangladesh.

\bibliographystyle{plain}
\bibliography{sample}
\end{document}